\UseRawInputEncoding
\documentclass[onecolumn]{aastex631}
\usepackage{float}
\usepackage{graphicx}
\pdfoutput=1%For ARXIV
\newcommand\gr{$\gamma$-ray}
\newcommand{\fermi}{\textit{Fermi}}
\newcommand{\phcms}{${\rm photons}\;{\rm cm}^{-2}\;{\rm s}^{-1}$} 
\usepackage{ulem}
\usepackage{xcolor}
\def\r{\textcolor{red}}
\def\b{\textcolor{blue}}
\def\p{\textcolor{purple}}

\graphicspath{{./}{figures/}}

\begin{document}

\title{Variability and Spectral Behavior of Gamma-Ray Flares of 3C 279}

\correspondingauthor{Gege Wang}
\email{wanggege@gzhu.edu.cn}

\correspondingauthor{Junhui Fan}
\email{fjh@gzhu.edu.cn}

\correspondingauthor{Hubing Xiao}
\email{hubing.xiao@shnu.edu.cn}

\author{Gege Wang}
\affiliation{Center for Astrophysics, Guangzhou University, Guangzhou 510006, China}
\affiliation{Astronomy Science and Technology Research Laboratory of Department of Education of Guangdong Province, Guangzhou 510006, China}

\author{Junhui Fan}
\affiliation{Center for Astrophysics, Guangzhou University, Guangzhou 510006, China}
\affiliation{Astronomy Science and Technology Research Laboratory of Department of Education of Guangdong Province, Guangzhou 510006, China}

\author{Hubing Xiao}
\affiliation{Shanghai Key Lab for Astrophysics, Shanghai Normal University, Shanghai, 200234, China}

\author{Jinting Cai}
\affiliation{Center for Astrophysics, Guangzhou University, Guangzhou 510006, China}
\affiliation{Astronomy Science and Technology Research Laboratory of Department of Education of Guangdong Province, Guangzhou 510006, China}

\begin{abstract}
3C 279 showed enhanced flux variations in \textit{Fermi}-LAT \gr\ observations from January to June 2018. 
We present a detailed \textit{Fermi}-LAT analysis to investigate the variability and spectral behaviors of 3C 279 during the \gr\ flares in 2018.
In this work, we analyzed the $\gamma$-ray spectra and found that the spectra in either the flaring or quiescent states do not show any clear breaks (or cutoffs).
This indicates that the dissipation region is outside the broad-line region, and the energy dissipation may be due to the inverse Compton process of scattering the dust torus infrared photons, this result is also consistent with that in \citet{Tolamatti2022}.
An external inverse Compton scattering of dusty torus (DT) photons is employed to calculate the broadband spectral energy distribution (SED).
This model was further supported by the fact that we found flare decay timescale was consistent with the cooling time of relativistic electrons through DT photons.
During the SED modeling, a relatively harder spectrum for the electron energy distribution (EED) is found and suggests these electrons may not be accelerated by the shock that happened in the dissipation region.
Besides, the magnetic reconnection is also ruled out due to a low magnetization ratio.
Thus, we suggest an injection of higher-energy electrons from outside the blob and raising the flare.
\end{abstract}


\section{Introduction} \label{sec:1}
It has been well established that blazars are the dominant \gr\ sources in the extragalactic sky \citep{3fagn15,4fagn20}. 
Blazars have a jet viewed under small angles closing to the line of sight. 
The jet orientation results in extreme observational properties, including multi-band rapid variations, high and variable polarization, strong and variable $\gamma$-ray emissions, and apparent superluminal motion, which are believed to be associated with a relativistic beaming effect \citep{Wills1992, Urry1995, Fan2002, Villata2006, Fan2014, Xiao2015, Gupta2016, Xiao2019, Abdollahi2020, Fan2021, Yang2022}.

Emission from the blazar jet dominates nearly over the entire electromagnetic spectrum and forms a two-bump structure spectral energy distribution (SED).
The lower energy bump is attributed to the synchrotron radiation of relativistic electrons.
The higher energy bump is attributed to the inverse Compton (IC) scattering, which is further divided into synchrotron-self Compton (SSC) if the soft photons come from the synchrotron process and external Compton (EC) if the soft photons come from external photon fields, e.g., the accretion disk \citep{Dermer1993}, the broad-line region (BLR) \citep{Sikora1994, Fan2006}, and the dusty torus (DT) \citep{Blazejowski2000, Arbeiter2002, Sokolov2005} in the frame of leptonic scenario.

3C 279 is a FSRQ with a black hole mass of (3$-$8)$\times10^{8}M_{\odot}$ \citep{nil+09} located at a redshift $z\sim0.536$ \citep{lyn+65}. 
It is known to be one of the bright and powerful \gr\ sources in the high-energy sky.
3C 279 was the first blazar showing strong and rapid variability at GeV energies detected by \textit{EGRET} onboard Compton Gamma Ray Observatory (CGRO) \citep{har+92}, also the first FSRQ detected above 100 GeV \citep{Magic08}. 
Since the launch of \textit{Fermi}-LAT in June 2008, the all-sky monitoring capability of the \textit{Fermi}-LAT has provided us with a continuous \gr\ flux investigation of 3C 279 for over 10 years. 
3C 279 showed several outbursts in the past 10 years, it went through a series of distinct flaring flux duration from 2013 December to 2014 April, hour-scale \gr\ flux variability was observed, with a maximum flux of $(6.54 \pm 0.30)\times10^{-6}\;$\phcms\ on April 03, 2014 \citep{pal+15}. 
On June 16, 2015, LAT observed a giant outburst from 3C 279 with a peak $>100$\,MeV flux of $\sim3.6\times10^{-5}\;{\rm photons}\;{\rm cm}^{-2}\;{\rm s}^{-1}$ 
averaged over orbital period intervals, with flux doubling time less than 5 min observed in its 2-min binned light curve \citep{ack+16}. 
It is the second FSRQ type blazar reported after PKS 1222+216 with a similar short minute timescale flux variation \citep{ale+11}. 
These studies provided constraints on the physical properties of the \gr\ emission region and emission processes \citep{Hayashida2015, Paliya2015}.

In 2018, 3C 279 became very active again with fluxes exceeding the previous 2015 level. 
Several studies have been carried out, mainly focusing on the correlation between multi-wavebands with hour binned light curves \citep{sha+19,pri+20,goy+22}. \citet{shu+20} analyzed the minute timescale peak-in-peak variability and proposed that the particle acceleration is due to relativistic magnetic reconnection.
However, the magnetic reconnection mechanism for the particle acceleration has been questioned for the reason of a low magnetization \citep{Hu2020, Hu2021, Tolamatti2022}, and the shock-in-jet model for the particle acceleration is considered for the 2018 flare \citep{Tolamatti2022}.

In the present work, we analyzed the LAT data in detail for the variability and spectral behaviors of 3C 279 during the \gr\ flares in 2018. Our measured \gr\ flux in 1-hour bins reached $\sim4.5\times10^{-5}\;$\phcms\ on April 19, 2018, and the source flux variability was resolved down to 2-min binned timescales, allowing an investigation of variability on minute timescales. The paper is organized as follows: In section~\ref{sec:2} we outline the \textit{Fermi}-LAT observations and the data analysis procedures. Following this, we calculate the broadband SED to investigate the main cause of the flux peak. We present the results of the discussion in the location of the \gr\ dissipation region and the jet kinematics during the flare in section~\ref{sec:3}, and a summary is given in section~\ref{sec:4}.

\section{\textit{Fermi}-LAT Gamma-Ray light curve and spectra analysis} \label{sec:2}
\subsection{Data analysis} \label{subsec:2.1}
LAT scans the whole sky every three hours in the energy range from 20 MeV to $>$ 300 GeV \citep{atw+09}. We selected LAT data from the Fermi Pass 8 database in the time period from 2008 August 4 15:43:36
(UTC) to 2020 Nov 25 23:41:00 (UTC), with energy range in 0.1--300 GeV. Following the recommendations of the LAT team\footnote{\footnotesize https://fermi.gsfc.nasa.gov/ssc/data/analysis/scitools/}, we selected events with zenith angles less than 90 deg to prevent possible contamination from the Earth's limb. The LAT science tool Fermitools 2.0.8 and instrument
response function (IRF) P8R3\_SOURCE\_V2 were used. For the target 3C 279, a 20$^\circ\times 20^\circ$ square region of interest (ROI) centered at its position given in 4FGL-DR2 \citep{4fgl+20} was selected. The normalization parameters and spectral indices of the sources within 5 deg from 3C 279, as well as sources within the ROI with variable index $\geq$ 72.44 \citep{ace+15}, were set as free parameters. All other parameters were fixed at their catalog values in 4FGL-DR2. We used the original spectral models in 4FGL-DR2 for the sources in the source model. Gamma-ray spectra and lightcurves binned by hour and minute timescales in subsection~\ref{subsec:2.3} were derived by an unbinned maximum likelihood analysis with \texttt{gtlike}. Galactic and extragalactic diffuse emission models were added to the source model using the spectral model file gll\_iem\_v07.fits and iso\_P8R3\_SOURCE\_V2\_v1.txt\footnote{\footnotesize https://fermi.gsfc.nasa.gov/ssc/data/access/lat/BackgroundModels.html}, respectively. The normalizations of the two diffuse emisson components were set as free parameters in the analysis.

\subsection{Long-term light curve} \label{subsec:2.2}
\begin{figure}
\epsscale{0.8}
\plotone{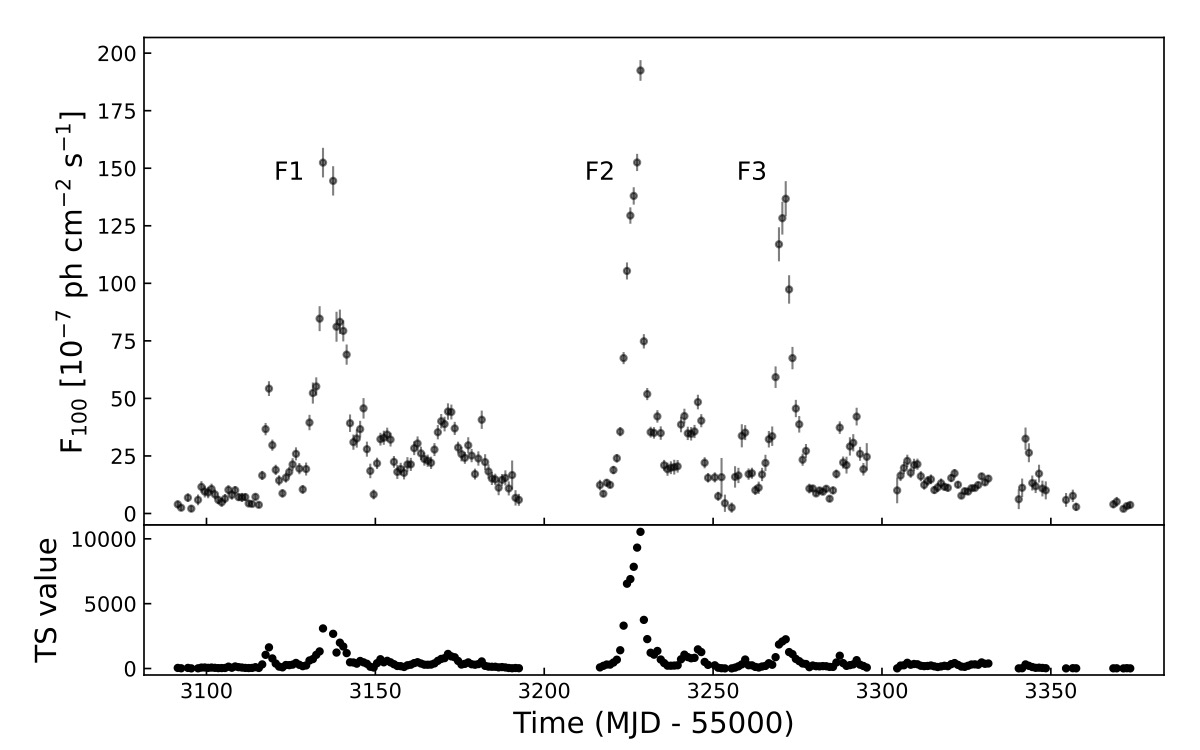}
\caption{Long-term LAT light curve and three pronounced \gr\ flares (F1, F2, F3) during MJD 58091 to MJD 58373 from 3C 279 with daily time bin in the energy range of 0.1--300 GeV.}
\label{fig:1dlc}
\end{figure}

We made the long-term light curve of the whole analysis time period above 100 MeV and found three obvious flares in 2018 in addition to the flare in 2015 June. Then we used the parameters in subsection~\ref{subsec:2.1} and a simple power law for 3C 279 in the source model. We constructed light curves in the daily time bin from MJD 58091 to MJD 58373 by performing a standard binned maximum likelihood analysis. The daily binned light curve is shown in Figure ~\ref{fig:1dlc}, in which when flux data points have the maximum likelihood Test Statistic (TS) values larger than 9 were plotted. The observed gap in Figure ~\ref{fig:1dlc} is due to a technical issue encountered on 2018 March 16 by \fermi\ spacecraft. LAT did not carry out any observations between March 16, 2018 and April 08, 2018.

\begin{figure}
\centering
\epsscale{0.75}
\plotone{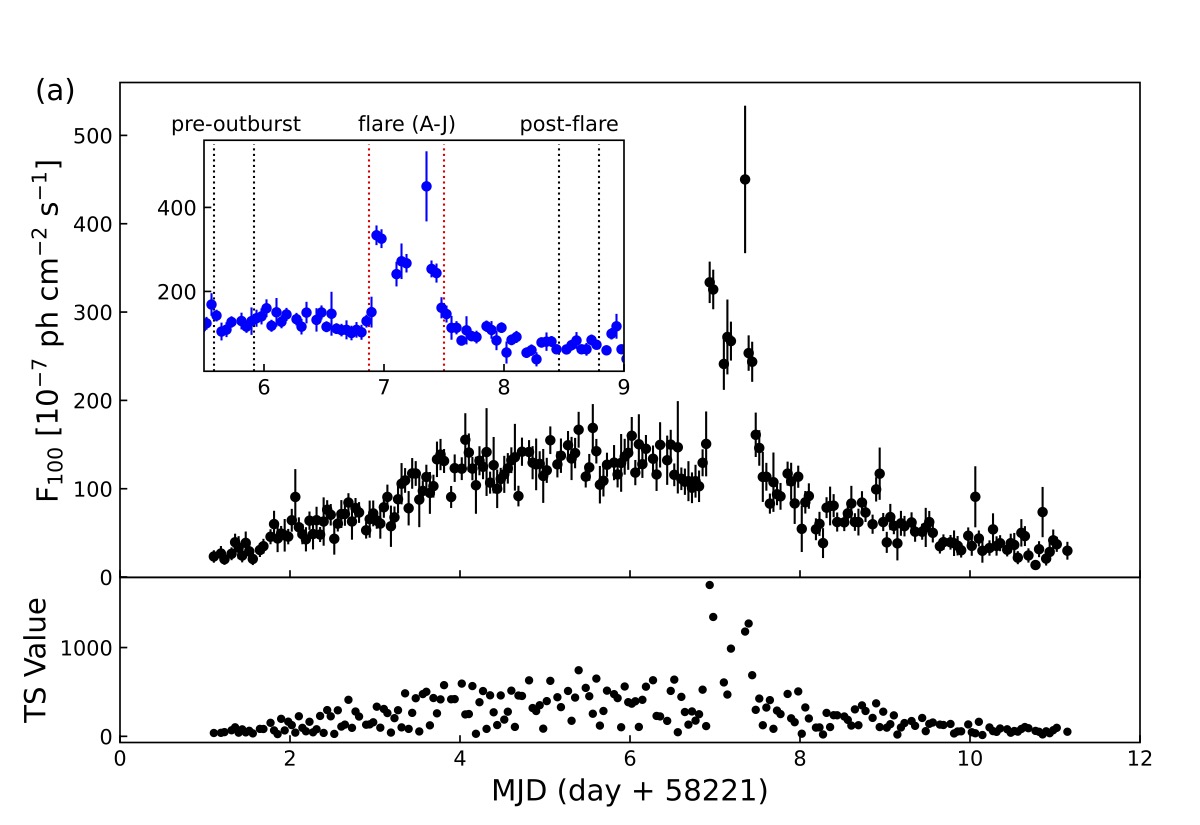}
\plotone{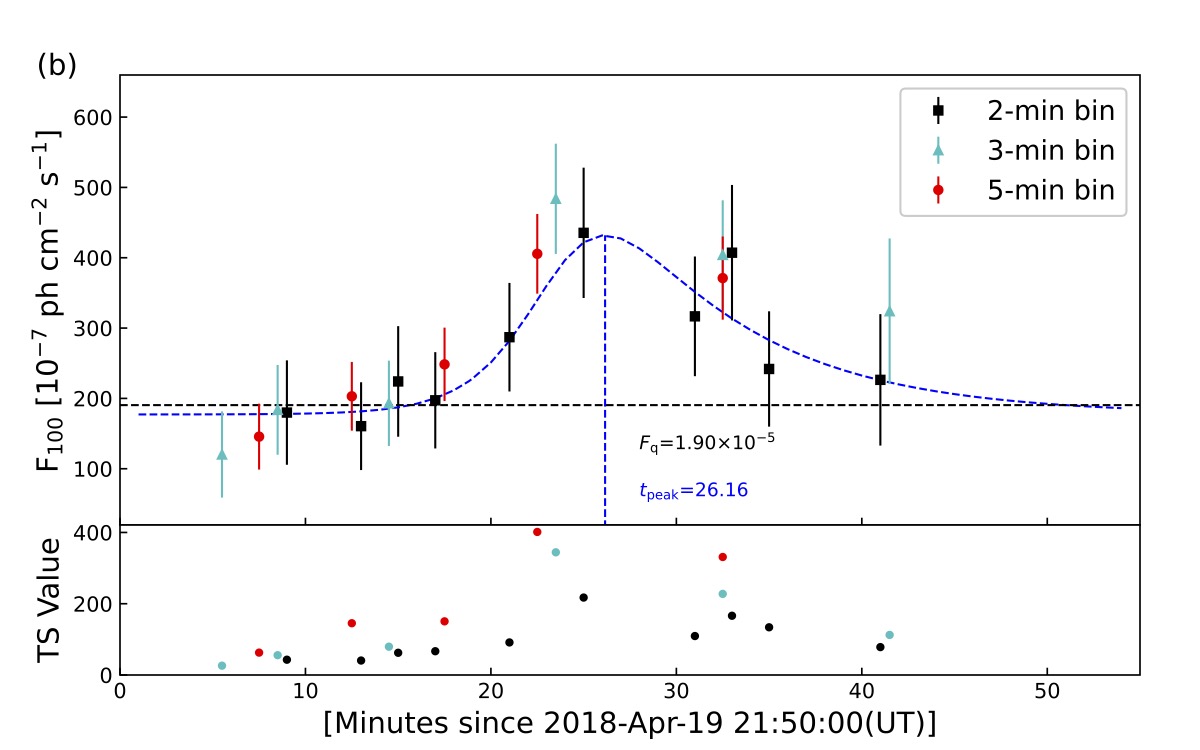}
\caption{(a) Shows 1-hour binned light curve during flare F2 of 3C 279 above 100 MeV. Blue points in the upper left panel show the enlarged view of the peak-in-peak variability. And we selected 10 bins as the flare period designated Stage "A" through "J", respectively. (b) Shows the 2-min (black square), 3-min (cyan triangle) and 5-min (red point) binned light curves measured during stage B of flare F2 wherein a strong rapid minute-scale variability was observed. During stage B, 3C 279 was found to be highly inconsistent with the constant flux having $p$-value $1.789\times10^{-15}$ ($\chi^{2}$-test) for the two-minute binned light curve. The best-fit function to the 2-min binned light curve is deduced using an exponential equation represented by the dashed blue curve. The black dashed line shows the flux in quiescence. }
\label{fig:flare}
\end{figure}

\subsection{Minute-scale flare} \label{subsec:2.3}

In order to investigate the fast flux variability in detail, we constructed light curves in hour timescale bins (Figure~\ref{fig:flare}-(a)), in which when flux data points have TS values larger than 9 were plotted. We noted that the flare F1 and F3 did not show noteworthy fast component or profile, therefore, we only made minute timescale binned light curve for F2 (Figure~\ref{fig:flare}-(b)). The measured \gr\ flux in 1-hour bins reached $\sim4.5\times10^{-5}\;$\phcms\ on April 19, 2018. Here we defined the flare period of F2 to be between MJD 58227.875 and MJD 58228.500 (April 19, 2018 21:00:00 and April 20, 2018 12:00:00), as indicated in Figure~\ref{fig:flare}-(a). And it concludes 10 bins designated Stage ``A'' through ``J'', respectively. We fitted a constant value to each stage in F2 for both time bins, and calculated a probability ($p$-value) from $\chi^{2}$ for each stage. While stage D, H, I resulted in $p$-values consistent with constant fluxes, we found significant indications of rapid variability on a minute timescale in the 2-min binned light curve. For stage B: ($p$, $\chi^2$/dof)$=$($1.789\times10^{-15}$, 87.11/8), stage C: ($p$, $\chi^2$/dof)$=$($3.370\times10^{-05}$, 47.24/15), stage E: ($p$, $\chi^2$/dof)$=$($1.747\times10^{-05}$, 31.85/6), stage F: ($p$, $\chi^2$/dof)$=$($2.145\times10^{-08}$, 51.45/8), stage G: ($p$, $\chi^2$/dof)$=$(0.013, 20.97/9).

We constructed light curves in 2-min, 3-min, and 5-min bins for the whole flare period of F2, and eventually found minute-scale flare profile only in stage B, which was shown in Figure~\ref{fig:flare}-(b). Then we used an exponential equation to fit the 2-min binned light curve profile, which is given by
\begin{equation}
F(t)=F_{\rm c}+F_{0}(e^{(t_{0}-t)/T_r} + e^{(t-t0)/T_d})^{-1}\ \ \ ,
\label{lcfiteq}
\end{equation}
where $F_{\rm c}$ and $F_{0}$ are the constant flux and height of a peak,
respectively, $t_{0}$ is the flux peak time, $T_r$ and $T_d$ are used to measure the rise and decay time separately. The best fitting result gives $\chi^2$/dof$=$1.71/5 with the flare peak time $t_{0}$ on 26.16 (2018 Apr 19 22:16:10(UTC)). Flare peak flux with the best fitting at $t_{0}$ is $F_{\rm p}=4.32\times10^{-5}\;$\phcms\ . We calculate the average flux value of the four data points before flare as the flux in quiescence ($F_{\rm q}=1.90\times10^{-5\;}$ \phcms\ ), which is shown in Figure~\ref{fig:flare}-(b). The time when the flux equals to $F_{\rm q}$ before and after the flare is 15.66 and 50.99, respectively. Therefore the flux decay time (from $F_{\rm p}$ to $F_{\rm q}$) is 24.83 min. We calculate the upper limit of $F_{\rm c}$ from the best fitting result of the exponential equation and take it as $F_{\rm q}$, which equals to $2.25\times10^{-5}\;$ \phcms\ . Then we can get that the lower limit of flux decay time is 15.04 min.

\begin{figure}
\centering
\epsscale{1.2}
\plotone{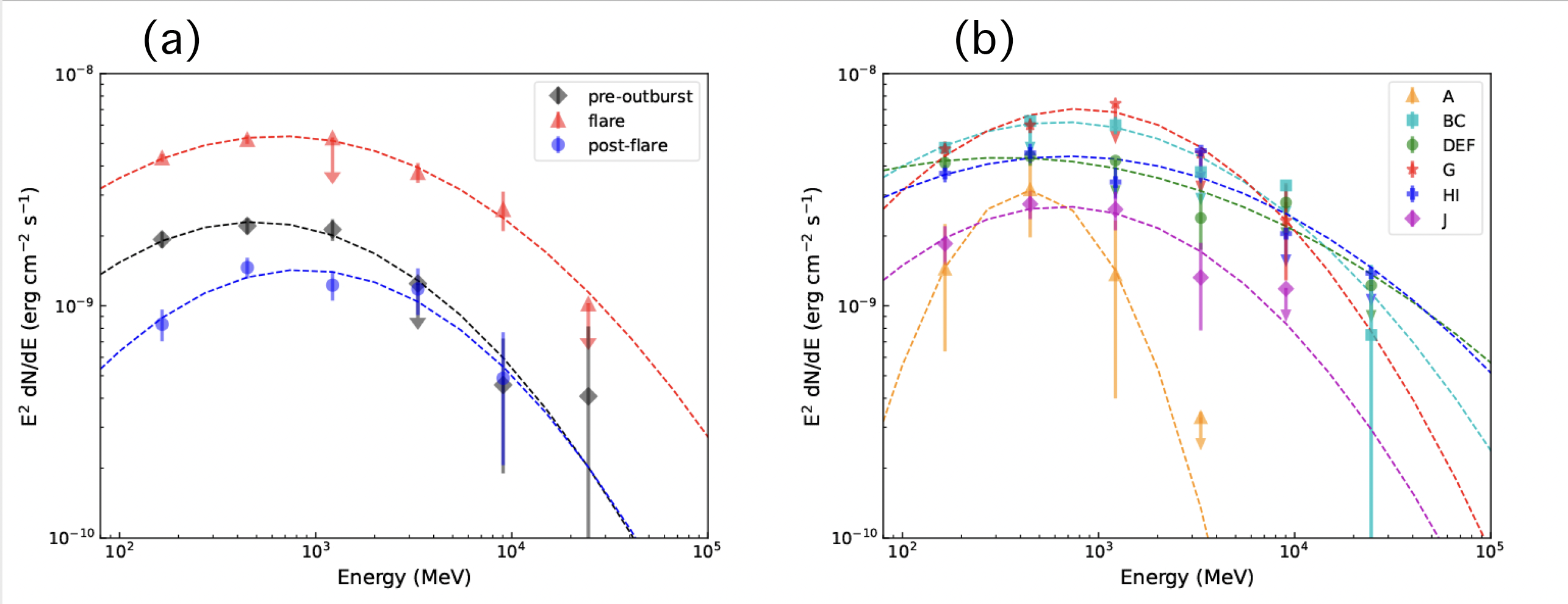}
\caption{Gamma-ray spectra of 3C 279 for each stage during the flare F2, as well as `pre-outburst' and `post-flare' as indicated in Figure ~\ref{fig:flare}. The downward arrows represent 95\% confidence level upper limits.}
\label{fig:spec}
\end{figure}

\subsection{The $\gamma$-ray spectral analysis} \label{subsec:2.4}
We fit the \gr\ spectra of 3C 279 described with a log-parabola model
($dN/dE \propto (E/E_0)^{-\alpha - \beta\log (E/E_0)}$, where
$\alpha$ and $\beta$ are spectral parameters, here $E_0$ = 466 MeV). We plotted the spectral data points with TS$>$4 and upper limits when TS$<$4 in Figure~\ref{fig:spec}. Figure~\ref{fig:spec}-(a) shows the spectra in the `{\it pre-outburst}' and `{\it post-flare}' periods as defined in Figure~\ref{fig:flare} for comparison. 
Figure~\ref{fig:spec}-(b) shows the \gr\ spectra as measured by \textit{Fermi}-LAT for each stage. Stages B and C, and stages D, E, and F, were combined because they showed similar fluxes and spectral fitting results. 
The spectral energy flux peaks in each stage correspond with the light curve during the flare F2. The flux and spectral fitting results above 100 MeV for Each stage (A−-J) during flare F2 are shown in Table~\ref{tab:table1}.

\begin{figure}
\centering
\epsscale{0.7}
\plotone{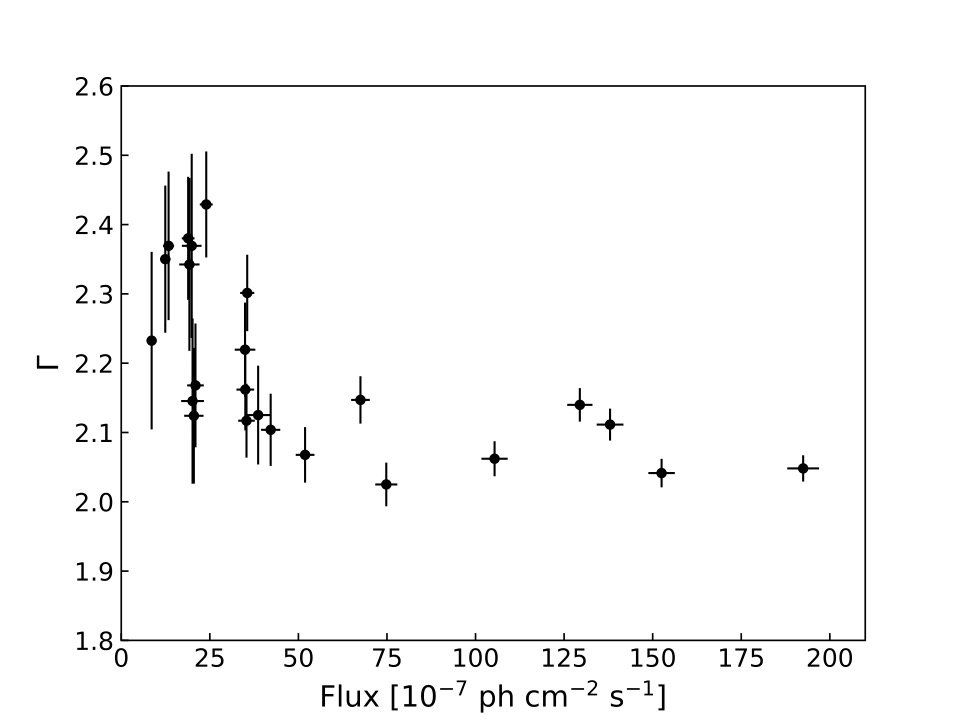}
\caption{Scatter plot of the \gr\ flux (0.1--300 GeV) vs. the \gr\ power law} photon index $\Gamma$ of 3C 279 with daily time bin based on the LAT data during the flare F2.
\label{fig:flux_index}
\end{figure}

The \gr\ flux (0.1--300 GeV) vs. the \gr\ power law photon index $\Gamma$ of 3C 279 with daily time bin based on the LAT data during the flare F2 is given in Figure ~\ref{fig:flux_index}. A correlation between the \gr\ flux and photon index was visible to suggest a `harder-when-brighter (BWB)' trend. 
The tendency of spectral hardening during the flaring state was seen in blazar flares by many authors (e.g., \citealt{bri+16,Shukla2018,sha+19}). 
A similar trend pattern was also previously seen for 3C 279 by \citet{Hayashida2015}, \citet{Paliya2015} and \citet{pri+20}.
The BWB could be explained in many ways, such as two-component (one stable component and a variable one) model \citep{Fiorucci2004}, fresh electrons injection \citep{Kirk1998, Mastichiadis2002}, etc.

\begin{deluxetable}{cccccl}
\label{tab:table1}
  \tabletypesize{\scriptsize} \tablecaption{Flux and Spectral Fitting Results of 3C 279 above 100 MeV for Each Stage (A-−J) during flare F2}
  \tablewidth{0pt}
    \tablehead{\colhead{Stage Number} & \colhead{Epoch (MJD)} &
 \colhead{Flux ($\times10^{-7}$)} & \colhead{$\alpha$} & \colhead{$\beta$} & \colhead{TS} } \startdata
pre-outburst & 58226.58$-$58226.92 &  123 $\pm$ 19   &  1.98 $\pm$ 0.06  &  0.16 $\pm$ 0.03  &     2598      \\
F2          &  58227.88$-$58228.50 &  270 $\pm$ 33  &   1.92 $\pm$ 0.03  &  0.12 $\pm$ 0.01   &   8156   \\
post-flare  &  58229.46$-$58229.79   &     67  $\pm$ 13   &  1.79 $\pm$ 0.12 &   0.17 $\pm$ 0.06   &   1451   \\ 
A       &   58227.83$-$58227.92     &    151 $\pm$ 37   &  2.07 $\pm$ 0.10  &  0.78 $\pm$ 0.12  &    115        \\
B$-$C      &      58227.92$-$58228.08    &    330 $\pm$ 23   &  1.92 $\pm$ 0.07 &   0.13 $\pm$ 0.03   &   2675    \\
D$-$F      &     58228.08$-$58228.21   &     260 $\pm$ 31  &   2.04 $\pm$ 0.08 &   0.06 $\pm$ 0.03   &   1553   \\
G        &     58228.21$-$58228.38    &    450 $\pm$ 43  &   1.80 $\pm$ 0.08 &   0.19 $\pm$ 0.04   &   1780    \\  
H$-$I      &      58228.38$-$58228.46   &     249 $\pm$ 21  &   1.93 $\pm$ 0.13 &   0.09 $\pm$ 0.05   &   1612   \\
J        &     58228.46$-$58228.54    &    161 $\pm$ 25  &   1.89 $\pm$ 0.12 &   0.17 $\pm$ 0.07   &   298    \\
\enddata
%\tablecomments{}
\end{deluxetable}

\section{Discussion} \label{sec:3}
\subsection{The location of the dissipation region}
The question of the location of the blazar $\gamma$-ray dissipation region ($d_{\rm diss}$) is intriguing to researchers.
To directly determine the $\gamma$-ray location, we need an extremely large $\gamma$-ray telescope to fulfill the required resolution at its wavelength.
Some indirect methods have been proposed to solve this issue, one of the popular ways is to measure the variability timescale and calculate the size of the dissipation region ($R_{\rm diss}$) by assuming the variability timescale accounts for the light travel across the dissipation region.
Timescale from hours to minutes have been reported for blazars \citep{abdo+11, Tanaka2011, Brown2013, Shukla2018} that constrain a very compact dissipation region.
Then, the size of the dissipation region is used to calculate distance to the central black hole through $d_{\rm diss}=R_{\rm diss}\psi$ with an assumption that the dissipation region diameter is identical to the width of the jet at this distance, where $\psi$ is the semi-aperture opening angle of the jet \citep{Dermer2009, Ghisellini2009}.
However, the distance is usually constrained to the base of the jet, where locates on the inner side of BLR, according to this method.
The BLR is full of optical/UV photons, which will cause absorption to the high energy $\gamma$-ray photons through pair production ($\gamma\gamma \to e^{\pm}$) \citep{Poutanen2010} that arises a spectral cut-off at GeV energies.
\citet{Acharyya2021} studied the $\gamma$-ray spectra for 18 flares and found that 11 of them with very high energy emission are incompatible with a BLR origin.

In the present work, we have studied the $\gamma$-ray spectra for the flare on April 19, 2018, see in Figure \ref{fig:spec}. 
Our results in Table \ref{tab:table1} indicate that a log-parabola model is enough to fit the spectra and no significant cut-off at GeV energies is observed during the flaring state, suggesting the dissipation region is outside the BLR and the energy dissipation may be due to the IC process of scattering the DT infrared photons.
This result is consistent with the one suggested by \citet{Tolamatti2022}.

\subsection{The broadband SED modeling}
We calculate the broadband SED for `pre-outburst', `flare', and `post-flare' states of F2.
We employ a log-parabolic-power-law (LPPL) function, which is a combination of a power-law function in the lower energy range and a log-parabola function at a higher energy range \citep{Tramacere2009, Tramacere2011, ta+20}, for the electron energy distribution (EED) in the dissipation blob \citep{Massaro2006}.
The LPPL function is expressed as follows:
\begin{equation}
N({\gamma})=\left\{
\begin{array}{llr}
N (\gamma / \gamma_{\rm 0})^{-s}  \ \ , \ \gamma \leq \gamma_{\rm 0} \\
\\
N (\gamma / \gamma_{\rm 0})^{-(s+r \cdot {\rm log}(\gamma / \gamma_{\rm 0}))}  \ \ , \ \gamma > \gamma_{\rm 0},
\end{array} \right.
\label{Eq_1}
\end{equation}
where $N$ is the normalization parameter, and $\gamma_{\rm 0}$ is the turn-over energy of the electron spectrum, $s$ is the spectral index, and $r$ is the spectral curvature.

In the framework of leptonic scenario \citep{Ghisellini1996, Tavecchio1998}, blazar broadband SED is formed through coupling the EED with the radiation mechanism (synchrotron radiation and inverse Compton scattering), and is usually displayed in the diagram of ${\rm log}\,\nu f_{\nu}-{\rm log}\, \nu$.
In the present work, we introduce the \textit{Jets SED modeler and fitting Tool} (\textit{JetSet}) \footnote{https://jetset.readthedocs.io/en/latest/index.html} \citep{Tramacere2009, Tramacere2011, ta+20} to generate SEDs.

There are nine basic parameters to describe the SED model.
Magnetic field ($B$), dissipation region size ($R_{\rm diss}$) and Doppler beaming factor ($\delta$) to describe the dissipation region environment and the jet geometry.
The rest of the six parameters, the maximum energy of electrons ($\gamma_{\rm max}$), the minimum energy of electrons ($\gamma_{\rm min}$), $\gamma_{\rm 0}$, $N$, $s$ and $r$ to describe the EED.
In the case of FSRQs, the entire $\gamma$-ray emission is a combination of self-Compton scattering (SSC), in which the soft photons are provided by the same population of relativistic electrons in synchrotron process, and external Compton scattering (EC), in which the soft photons can have multi-suppliers e.g., accretion disk \citep{Dermer1993}, the broad emission line region (BLR) \citep{Sikora1994, Fan2006}, and the dusty torus (DT) \citep{Blazejowski2000, Arbeiter2002, Sokolov2005}.

%this is for DT model

We consider the EC model with soft photons origin from the DT, and there are 5 additional parameters to define the accretion disk and the DT, the disk luminosity ($L_{\rm disk}$), the disk temperature ($T_{\rm disk}$), the distance of the DT from the central engine ($R_{\rm DT}$), the temperature of the DT ($T_{\rm DT}$), and the opacity of the DT ($\tau_{\rm DT}$).

The simultaneously observed data that we used for SED calculation are collected from \citet{Tolamatti2022}, in which the multi-wavelength data of duration MJD 58218--58243 is concluded.
This duration spans 25 days and has a full coverage of flare F2.
Thus, we collect radio, optical/UV, and X-ray data from their work to combine with our GeV $\gamma$-ray data of the three states.
During the SED fitting, we have six parameters fixed as tabulated in Table \ref{tab_fix} and the rest seven parameters to vary in Table \ref{tab_unfix}, and the quantities related to jet power and SED structure for all the three states are listed in Table \ref{tab_result} .

%this is for DT model
\begin{deluxetable*}{lc}
\label{tab_fix}
\renewcommand{\arraystretch}{1.3}
\tabletypesize{\scriptsize} \tablecaption{Fixed parameters in the broadband SED modeling of 3C 279 flare F2}
%\tablewidth{0pt}
\tablehead{
\colhead{Parameter} & 
\colhead{Fixed value for all states} 
} 
\startdata
$T_{\rm Disk} \ {\rm (K)}$                          &   $2 \times 10^{4}$       \\
$L_{\rm Disk} \ {\rm (erg \cdot s^{-1})}$           &   $2 \times 10^{45}$      \\
$R_{\rm diss} \ {\rm (cm)}$                                    &   $7 \times 10^{16}$      \\
$R_{\rm H} \ {\rm (cm)}$                            &   $1 \times 10^{17}$      \\
$\gamma_{\rm max}$                                  &   $1 \times 10^{6}$       \\
$\tau_{\rm DT}$                                     &   0.15                    \\
\enddata
\end{deluxetable*}

%this is for DT model
\begin{deluxetable*}{lccc}
\label{tab_unfix}
\tabletypesize{\scriptsize} \tablecaption{Parameters in the broadband SED modeling of 3C 279 flare F2}
\tablewidth{0pt}
\tablehead{
\colhead{Parameter} & 
\colhead{`pre-outburst' state} &
\colhead{`flare' state} &
\colhead{`post-flare' state} 
} 
\startdata
$B \ {\rm (G)}$            &   0.13    &    0.12   &    0.13   \\
$\delta$                    &   32.3    &    40     &    28     \\
$N \ {\rm (cm^{-3})}$       &   230     &    240    &    250    \\
$\gamma_{\rm 0}$            &   220     &    150    &    280    \\
$\gamma_{\rm min}$          &   20      &    10     &    20     \\
$s$                         &   1.81    &    1.65   &    1.81   \\
$r$                         &   0.67    &    0.69   &    0.64   \\
\enddata
%\tablecomments{}
\end{deluxetable*}

\begin{deluxetable*}{lccc}
\label{tab_result}
\tabletypesize{\scriptsize} \tablecaption{SED structure and jet power quantities obtained through the modeling}
\tablewidth{0pt}
\tablehead{
\colhead{Parameter} & 
\colhead{`pre-outburst' state} &
\colhead{`flare' state} &
\colhead{`post-flare' state} 
} 
\startdata
$\nu_{\rm syn} \ {\rm (Hz, log)}$       &   13.45    &    13.29   &    13.69    \\
$\nu_{\rm IC} \ {\rm (Hz, log)}$        &   23.23    &    23.23   &    23.43    \\
$\nu_{\rm DT} \ {\rm (Hz, log)}$        &   13.57    &    13.57   &    13.57    \\
$P_{\rm e} \ {\rm (erg \cdot s^{-1})}$  &   $9.59 \times 10^{45}$   &   $1.09 \times 10^{46}$   &   $8.55 \times 10^{45}$   \\
$P_{\rm p} \ {\rm (erg \cdot s^{-1})}$  &   $1.66 \times 10^{46}$   &   $2.66 \times 10^{46}$   &   $1.36 \times 10^{46}$   \\
$P_{\rm B} \ {\rm (erg \cdot s^{-1})}$  &   $3.23 \times 10^{44}$   &   $4.23 \times 10^{44}$   &   $2.43 \times 10^{44}$   \\
$P_{\rm r} \ {\rm (erg \cdot s^{-1})}$  &   $3.76 \times 10^{45}$   &   $5.45 \times 10^{45}$   &   $3.24 \times 10^{45}$   \\
\enddata
%\tablecomments{}
\end{deluxetable*}

%this is for DT model

\begin{figure}
\centering
\epsscale{0.7}
\plotone{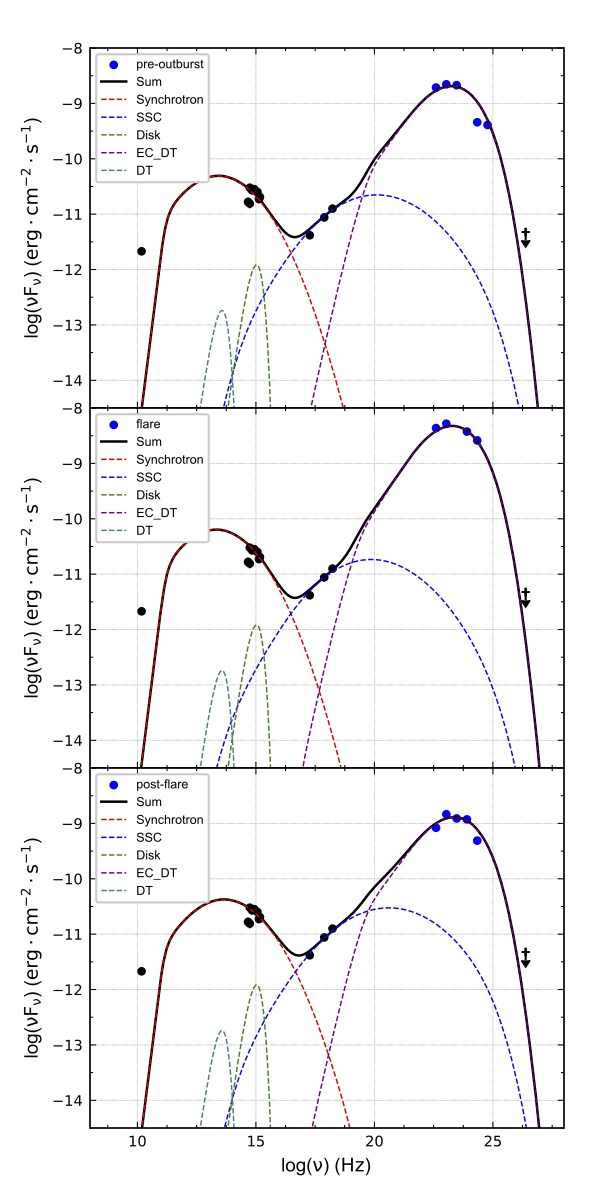}
\caption{SEDs of the `pre-outburst', the `flare', and the `post-flare' states.}
\label{fig:figures}
\end{figure}

\subsection{The jet kinematics}
The blazar, 3C 279 at the redshift 0.536, is one of the extreme variable celestial objects in multi-wavelengths.
In 2018, it demonstrated three outbursts, namely F1, F2, and F3, with GeV $\gamma$-ray fluxes exceeding the previous flare in 2015.
Especially, an unprecedented high flux was observed in flare F2.

We, in particular, analyzed the spectral behavior of flare F2, which pronounced `pre-outburst', `flare' and `post-flare', seen in Figure \ref{fig:spec}.
The spectral fitting results above 100 MeV are shown in Table \ref{tab:table1}.
It showed clearly that a `harder-when-brighter' trend was observed for F2, which suggests the detection of a high-energy photon during the high state.

During the broadband SED modeling, we fixed parameters and used those values as in \citet{Tolamatti2022}.
In their work, those parameters were optimized during the fitting procedure and consistent with those values in literature \citep{Hayashida2015, Paliya2015}.
In addition, we also assumed a fraction $\tau_{\rm DT}=0.15$ of the disk luminosity was intercepted and re-radiated from dust as IR emission, as a comparison, this value was used as 0.5 for \citet{Aleksic2011} and in \citet{Paliya2015} and was used as 0.1 in \citet{Hayashida2012, Hayashida2015}.

One should be able to calculate the jet power with these given fixed parameters and those parameters obtained through SED modeling.
The jet power $P_{\rm jet}$, which is a critical parameter to understand the formation and composition of jets, is usually assumed to be carried by relativistic electrons ($P_{\rm e}$), magnetic field ($P_{\rm B}$), radiation ($P_{\rm r}$) and cold proton kinetic ($P_{\rm p}$) \citep{Celotti2008, Tan2020} as:
\begin{equation}
P_{\rm jet} = \Sigma_{i} \pi R^{2} \Gamma^{2} c U_{i}, 
\label{Pjet}
\end{equation}
where $U_{i}$ represents the energy density of the magnetic field ($i=B$), the relativistic electron ($i=e$), cold proton ($i=p$) and bolometric radiation ($i=r$) \citep{Ghisellini2010, Ding2019} in the comoving frame, and can be estimated in \citet{Tan2020}:
\begin{equation}
U_{\rm e} = m_{\rm e}c^{2} \int N(\gamma)\gamma d\gamma, 
\label{Ue}
\end{equation}

\begin{equation}
U_{\rm p} = m_{\rm p}c^{2} \int 0.1 N(\gamma) d\gamma, 
\label{Up}
\end{equation}
where we applied a standard assumption of one cold proton every ten electrons;

\begin{equation}
U_{\rm B} = B^{2}/8\pi, 
\label{UB}
\end{equation}

\begin{equation}
U_{\rm r} = L_{\rm obs}/(4\pi R^{2}c\delta^{2}), 
\label{Ur}
\end{equation}
where $L_{\rm obs}$ is the observed non-thermal bolometric luminosity, which is estimated by integrating over the broadband SED.
The $P_{\rm e}$, $P_{\rm p}$, $P_{\rm B}$, $P_{\rm r}$ are calculated and tabulated in Table \ref{tab_unfix}.

Our results suggest the magnetic field stays stable, but we find a slightly higher magnetic field for low activity states and a lower magnetic field for the high emission state, which has also been reported in previous works \citep{sha+19, pri+20}.
Besides, the Doppler factor varies and the relativistic electron distribution changed significantly during the `pre-outburst', `flare', and `post-flare' states.
The Doppler factor gives the largest value for the `flare' state, while giving the smallest value for the `post-flare' state among the three states;
This is consistent with the fact found in \citet{sha+19}, in which they studied the flare of 3C 279 in 2018 January and found that the bulk Lorentz factor increased when the flux increased.
During the `flare' state, the electron energy distribution is significantly changed, $\gamma_{\rm min}$, $\gamma_{\rm 0}$, and $s$ are becoming smaller, compared to those of the `pre-outburst' state and the `post-flare' state.

Different explanations of this flare have been purposed by previous studies.
\citet{shu+20} analyzed the light curve of F2 and found a $\sim$8 minute-timescale flare, which was superimposed on a longer duration envelope, and a peak-in-peak structure in this flare.
They suggested a magnetic reconnection-derived particle acceleration to form the fast flare as well as the observed peak-in-peak light curve structure.
However, \citet{Tolamatti2022} suggested that the flare should not be caused by a magnetic reconnection process, because they found the jet is particle dominated ($U_{\rm e}/U_{\rm B} \gg 1$) and the magnetization ratio ($\sigma_{\rm B} = P_{\rm B}/(P_{\rm e} + P_{\rm p})$) is too low to accelerate particles via magnetic reconnection, base on what they suggested the shock-in-jet may still responsible for the particle acceleration in the jet.

In analogy, we calculate the $U_{\rm e}/U_{\rm B}$ and $\sigma_{\rm B}$ and report them in Table \ref{tab_par}, our results suggest that the particle energy density dominates over magnetic energy density.
And the low values of $\sigma_{\rm B} \sim 0.01$ indicate that magnetic reconnection is less likely to account for particle acceleration, but it can accelerate electrons via relativistic shocks as suggested by \citet{Baring2017}, in which electrons would be efficiently accelerated by relativistic shocks in blazar jets with $\sigma_{\rm B}$ ranges from $\sim 10^{-4}$ to 0.06.

%for the DT model
\begin{deluxetable*}{lccc}
\label{tab_par}
\tabletypesize{\scriptsize} \tablecaption{The magnetization ratio and the electron to magnetic field energy density ratio}
\tablewidth{0pt}
\tablehead{
\colhead{Parameter} & 
\colhead{`pre-outburst' state} &
\colhead{`flare' state} &
\colhead{`post-flare' state} 
} 
\startdata
$\sigma_{\rm B}$            &    0.012   &   0.011    &    0.011   \\
$U_{\rm e}/U_{\rm B}$       &    29.61   &   25.65    &    35.12   \\
\enddata
%\tablecomments{}
\end{deluxetable*}

Our results of Table \ref{tab_par} meet the conclusion from \citet{Tolamatti2022}, but the contradiction between the shock acceleration in the emission blob and the minute-scale variability still exist that mentioned by \citet{shu+20}.
Besides, the EED with a spectral index $\sim 2$ predicted by the non-relativistic shock acceleration and $\sim 2.2$ for the case of a classic relativistic shock \citep{Baring1999, Kirk2000, Ellison2004, Hu2020}, but we obtained a much harder electron spectrum with $s=1.65$ for the flaring state.

Instead of both the shock acceleration and the magnetic reconnection, we consider this flare is possibly caused by an injection of higher-energy electrons from outside of the blob, these electrons are accelerated to higher energy outside and injected into the blob in a very short time.
This injection of higher-energy electrons changes the distribution of electrons to a harder spectrum and the increases amount of electrons with energy above $\gamma_{\rm 0}$.
These higher-energy electrons could be accelerated by a more comprehensive mechanism that can explain the unique light curve behaviour (i.e., `peak-in-peak' structure), for instance, they are accelerated by the magnetic reconnection outside the blob and also obtain energy from the shock inside the blob.
Otherwise, a multi-zone model should be involved to explain all these phenomenon.

\subsection{The cooling of relativistic electrons}
In the leptonic scenario, the jet power is mostly dominated by the kinetic power of particles, including the relativistic electrons (or positrons) and cold protons.
Energy dissipation of relativistic electrons is more efficient through the IC process than the synchrotron radiation resulting in a $\gamma$-ray dominant emission from the jet.
The observed spectra (see in Figure \ref{fig:spec} and Table \ref{tab:table1}) of the flare state favor a model of EC\_DT, and the broadband SED has been constructed in the previous subsection.
The cooling time of electrons is able to be estimated in the observer frame through
\begin{equation}
t^{\rm ob}_{\rm cool}=\frac{3m_{\rm e}c(1+z)}{4\sigma_{\rm T}\gamma \delta U^{'}},
\end{equation}
where $m_{\rm e}$ is the electron rest mass, $\gamma$ is the electron energy in unit of Lorentz factor, and the $U^{'}$ is the energy density in the jet frame.
The Lorentz factor $\gamma$ can be derived by the IC peak frequency: $\nu_{\rm IC} \approx \nu_{\rm ext}\gamma^{2} \Gamma \delta/(1+z)$, then
\begin{equation}
t^{\rm ob}_{\rm cool}=\frac{3m_{\rm e}c}{4\sigma_{\rm T} U^{'}} \left(\frac{\nu_{\rm ext}}{\nu_{\rm IC}}\right)^{1/2} \left(\frac{\Gamma}{\delta}\right)^{1/2} (1+z)^{1/2},
\end{equation}
the external photon energy density $U^{'}=3 \times 10^{-4}\, \Gamma^{2} \ {\rm erg \cdot cm^{-3}}$ and the bulk Lorentz factor ($\Gamma$) is believed to be close to the Doppler factor ($\delta$) for blazars due to a small viewing angle \citep{Ghisellini2008, Tavecchio2010, Foschini2011}.
We calculate the cooling time scale for our EC\_DT model and get $t^{\rm ob}_{\rm cool}= 30.1 \, {\rm min}, \, 19.6 \, {\rm min}, \, 31.8 \, {\rm min}$ for the three states, respectively.
An average cooling time scale $t^{\rm ob}_{\rm cool}= 27.2 \pm 6.6 \, {\rm min}$, which meets the flux decay time scale that we obtained in the previous section (see also in Figure \ref{fig:flare}).
The result suggests that the flux decay is dominated by the EC cooling of scattering DT photons and proves the dissipation region is located outside the BLR.

\section{Summary} \label{sec:4}

In this paper, we analyze the LAT data and investigate the rapid variability and spectral behaviors of quasar 3C 279 during the three \gr\ flares in 2018. In the obtained daily binned long-term LAT light curve, we pronounced three \gr\ flares (F1, F2, F3) during MJD 58091 to MJD 58373 in the energy range of 0.1--300 GeV. We noted that the flares F1 and F3 did not show noteworthy fast components or profiles, therefore, we only made a detailed minute timescale binned light curve and a \gr\ spectral analysis for F2. The \gr\ flux peak in F2 has exceeded the previous flare level in 2015. Then we made the broadband SED modeling, calculated the jet kinematic parameters and the cooling time of relativistic electrons, discussed the location of the \gr\ dissipation region. and came to the following conclusions:

(i) Our \gr\ spectra fitting results indicate that a log-parabola model is enough to fit the spectra during the flaring state in F2. This suggests the dissipation region is outside the BLR and the energy dissipation may be due to the IC process of scattering the DT infrared photons. 

(ii) We calculate the broadband SED for `pre-outburst', `flare', and `post-flare' states of the flare F2. We consider the EC model with soft photons origin from the DT and use the simultaneously observed multi-wavelength data from MJD 58218 to MJD 58243 for SED calculation. 

(iii) We calculate the magnetization ratio and the electron to magnetic field energy density ratio. According to the results, we consider the flare F2 is possibly caused by an injection of higher-energy electrons from outside of the blob. Electrons are accelerated to higher energy outside and injected into the blob in a very short time. 

(iv) Our calculation result of the cooling time scale for our EC\_DT model meets the flux decay time in the LAT observation. This result suggests that the flux decay is dominated by the EC cooling of scattering DT photons, and also proves the dissipation region is located outside the BLR.

\begin{acknowledgments}
Thanks are given to the reviewer for the constructive comments and helpful suggestions. We thank Dr. A. Tolamatti for sharing the broadband SED data with us. The work is partially supported by the National Natural Science Foundation of China (NSFC U2031201, NSFC 11733001), Guangdong Major Project of Basic and Applied Basic Research (Grant No. 2019B030302001). 

We also acknowledge the science research grants from the China Manned Space Project with NO. CMS-CSST-2021-A06, Scientific and Technological Cooperation Projects (2020-2023) between the People's Republic of China and the Republic of Bulgaria.

\end{acknowledgments}

\bibliography{3c279flare}
\bibliographystyle{aasjournal}

\end{document}